# Multimedia stimuli databases usage patterns: a survey report

M. Horvat[1], S. Popović[1] and K. Ćosić[1]
[1] University of Zagreb, Faculty of Electrical Engineering and Computing
Department of Electric Machines, Drives and Automation
Zagreb, Croatia
marko.horvat2@fer.hr

**Abstract** – Multimedia documents such as images, sounds or videos can be used to elicit emotional responses in exposed human subjects. These stimuli are stored in affective multimedia databases and successfully used for a wide variety of research in affective computing, human-computer interaction and cognitive sciences. Affective multimedia databases are simple repositories of multimedia documents with annotated high-level semantics and affective content. Although important all affective multimedia databases have numerous deficiencies which impair their applicability. To establish a better understanding of how experts use affective multimedia databases an online survey was conducted into the subject. The survey results are statistically significant and indicate that contemporary databases lack stimuli with rich semantic and emotional content. 73.33% of survey participants find the databases lacking at least some important semantic or emotion content. Most of the participants consider stimuli descriptions to be inadequate. Overall, 1-2h or more than 24h are generally needed to construct a single stimulation sequence. Almost 84% of the survey participants would like to use real-life videos in their research. Experts unequivocally recognize the need for an intelligent stimuli retrieval application that would assist them in experimentation. Almost all experts agree such applications could be useful in their work.

## I. Introduction

Emotionally annotated databases such as International Affective Picture System (IAPS) [1], International Affective Digital Sounds (IADS) [2], The Geneva Affective PicturE Database (GAPED) [3] and NimStim Face Stimulus Set [4] are often used in research of emotion processing, attention, stress resilience and mental health, yet not enough has been done to facilitate their usage and expand prevalence in the field. Multimedia documents are loosely annotated which makes semantic retrieval difficult and results in low recall and precision. High-level semantic content descriptors are informal and ambiguous due to the insufficient annotation methods which rely on unrestricted keywords. In this limited meta-retrieval framework no knowledge base, concept taxonomy or terminology exists. These problems contribute to a deficient stimulation, suboptimal elicitation and emphasize the need to thoroughly improve the elicitation process and the structure of multimedia stimuli databases [5] [6].

Currently stimuli from multimedia databases are extracted manually by a laborious visual examination of each stimulus and the accompanying manuals. Stimuli sequence construction is often demanding and time-consuming task.

To complicate their usage even further, multimedia stimuli databases have diverse structures, describe emotional and semantic data differently and contain various media formats and stimuli modalities [5]. There is no consensus among researchers in an optimal format or implementation of a multimedia stimulus database [6]. Because of these reasons a typical multimedia stimuli database user must be a proficient expert in the field of emotion elicitation and simultaneously skilled in technology-related tasks such as stimuli selection and extraction. Such heterogeneous fusion of different skills is difficult to master and since databases' structures are mutually different these skills must be separately trained for each multimedia stimuli database [7].

All these issues outline a need for a stimuli generating computer system which can assist experts in finding the most appropriate stimulus and to do it in the shortest time possible. Such system must be as universal as possible towards various database and media formats, efficient and user-friendly. Furthermore, to assist the expert even further the system should contain empirically derived rules for decision support and automatic generation of stimuli sequences. Also, the stimuli generator should support personalized stimulation individually tailored to specific emotional and semantic parameters.

To address these issues, and as a part of research in methods for formal representation of multimedia stimuli, a software application Intelligent Stimuli Generator (intStimGen) was developed at the Laboratory for Interactive Simulation Systems with the University of Zagreb, Faculty of Electrical Engineering and Computing. The intStimGen (Fig. 1) enables searching of multimedia stimuli databases and construction of stimuli sequences using semantic and emotion descriptors. Stimuli semantics can be described with unrestricted keywords, tag clouds, WordNet synsets [8] and SUMO concepts [9], while emotion descriptors are based on dimensional models. Integrated exploration of IAPS and IADS is supported, but the application's architecture enables other stimuli databases to be modularly added in the future.

This research has been partially supported by the Croatian Ministry of Science, Education and Sports.

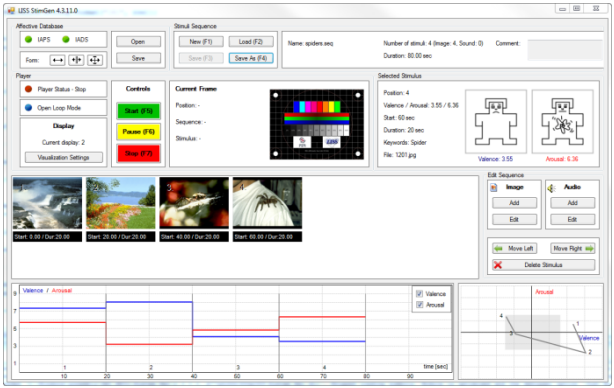

Figure 1. The main screen of the Intelligent Stimuli Generator application displaying a fear-provoking sequence constructed with semantics related to arachnophobia.

The constructed sequences can be displayed to a subject on a separate screen and his voluntary responses may be acquired through a standard human interface device such as keyboard. During exposure the application sends system messages which can be used for synchronization of physiology acquisition equipment and other systems.

However, to assess the need for such software system among the professional community a survey was needed to collect experiences and opinions among experts who use multimedia stimuli in their research.

## II. SURVEY METHOD

Between 19 March and 19 April 2012 an invitation to a survey has been sent to 120 e-mail addresses of different researchers, psychologists, psychiatrists, neuroscientists and medical doctors that have been published papers in which they have used at least one emotionally annotated database. The targeted databases primarily were IAPS, IADS[2], NimStim and GAPED. Some invitees have been using Karolinska Directed Emotional Faces (KDEF) [10], Pictures of Facial Affect (POFA) [11] and Japanese Female Facial Expression (JAFFE) Database [12]. Relevant papers were retrieved using search engines Web of Knowledge, PubMed, IEEE Xplore and Google Scholar. The invitees were all outside Croatia (i.e. international) and their affiliations included Max Planck Institute for Human Cognitive and Brain Sciences, Georg-August-University Göttingen, Humboldt-Universität zu Berlin, University of Geneva, University of Gent, University of Chicago, University of Florida, University of Southern California etc. The participants could fill out a survey form in Word format or give their answers online at: http://www.surveymonkey.com/s/L79ZXV2. Out of 120 invitees 30 have completed the survey (25%). Four (4) have filled out the survey form and other 26 have completed the survey online. The participants' answers were sent in period 26 March – 28 August 2012. The answers were individually recorded and archived. The anonymity of the participants was guaranteed as was stated in the e-mail invitation. Only participants' IP addresses, the time of access and responses were recorded. Multiple answers were not allowed. Online participants finished the survey in under 5 minutes, with 2min 40sec on average.

The survey consisted of 10 question with predefined answers. Possible answers in the Likert scale were in a range 1–5 or 1–7. All questions also had "Not sure / Not applicable" ("N/A") option as one of the possible answers. Any question could be left unanswered (blank). Such responses were later processed as "N/A".

The survey questions were:

1. How difficult or easy is the image retrieval process from an emotionally-annotated database (e.g. IAPS, NimStim, IAD), KDEF, POFA, JAFFE, GAPED, etc.)?

2. How satisfied are you with the above mentioned level of difficulty in the image retrieval process from the database?

3. How much time was necessary to effectively search the database and construct one picture sequence that was used in your research?

4. Have you at any time felt that the picture set you were using is missing images with particular emotion or semantic content that would be useful for your stimuli sequence?

5. On a scale of 1 to 7, with one being extremely inadequate and 7 being extremely adequate, please rate how inadequate, ambiguous or insufficient did you find the predefined descriptions of the images you used?

6. On a scale 1 to 7, with one being extremely useless and 7 being extremely useful, please rate how useful, helpful or beneficial would a user-friendly software tool for intelligent retrieval of emotionally-annotated images be to your research?

7. Have you at any time during your research wanted to find the most appropriate emotionally-annotate image faster and more efficiently?

8. Did you construct the sequence manually or with a help of a software tool?

9. Skip this question if you did not use a software tool for retrieval of emotionally-annotated images. Did your group actually develop the tool used in your experiment or acquired it elsewhere?

10. How useful or useless a stimuli database with real-life video-clips, instead of just still images, would be to your work?

## III. CUMULATIVE SURVEY RESULTS

The survey results are given in the table and figure below. The results are freely available and can be acquired by contacting the first author of the paper (i.e. the author of the survey).

TABLE I. OVERVIEW OF THE CUMULATIVE SURVEY RESULTS

| Question[1] | Answers |
|---|---|
| Q1 | Very difficult = 0%, Difficult = 16.67%, Neither difficult nor easy = 36.67%, Easy =20%, Very easy = 10%, N/A = 16.67% |
| Q2 | Very dissatisfied = 3.33%, Dissatisfied = 13.33%, Neither satisfied nor dissatisfied = 33.33%, Satisfied = 30%, Very satisfied = 6.67%, N/A = 13.33% |
| Q3 | Less than 5 mins = 0%, 5-15 mins = 3.33%, 15-30 mins = 10%, 30-60 mins = 10%, 1-2 hrs = 26.67%, 2-6 hrs = 13.33%, 6-12 hrs = 10%, More than 12 hrs = 20%, N/A = 6.67% |
| Q4 | Yes = 73.33%, No = 20%, N/A = 6.67% |
| Q5 | Value 1 (extremely inadequate) = 0%, Value 2 = 6.67%, Value 3 = 30%, Value 4 (neutral) = 23.33%, Value 5 = 23.33%, Value 6 = 10%, Value 7 (extremely adequate) = 0%, N/A = 6.67% |
| Q6 | Value 1 (extremely useless) = 0%, Value 2 = 0%, Value 3 = 0%, Value 4 (neutral) = 10%, Value 5 = 23.33%, Value 6 = 13.33%, Value 7 (extremely useful) = 46.67%, N/A = 6.67% |
| Q7 | Yes = 86.67%, No = 10%, N/A = 3.33% |
| Q8 | Manually = 46.67%, With a software tool = 23.33%, Both = 23.33%, N/A = 6.67% |
| Q9 | The tool was acquired = 20%, The tool is own development = 20%, N/A = 60% |
| Q10 | Completely useless = 0%, A little bit useful = 3.33%, Neither useful nor useless = 0%, Useful = 20% Very useful = 73.33%, N/A = 3.33% |

1. Text of the questions is provided in Section 2.

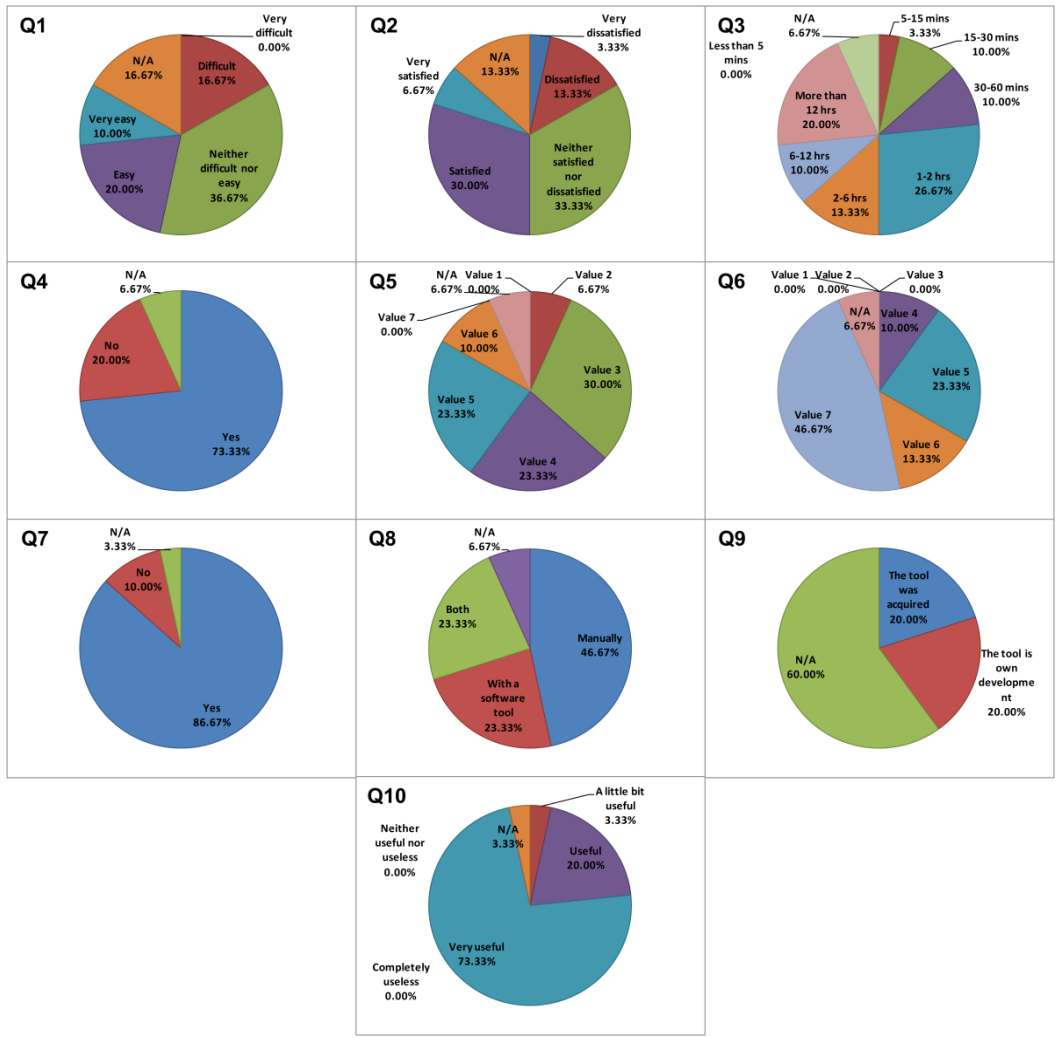

Figure 2. Corresponding charts of the cumulative survey results.

## IV. DISCUSSION AND RESULTS ANALYSIS

The survey has clearly indicated that intelligent, efficient and user-friendly generators of emotionally and semantically annotated multimedia stimuli are clearly desired by the professionally community. A great majority of the participating researchers (83.33%) consider such intelligent software tools to be useful or very useful in their work with retrieval of emotionally-annotated images.

Very importantly the results show that stimuli sequences are predominantly constructed by hand – with experts thoroughly scouring through multimedia stimuli– over a period of 1-2 hours (26.67%), 2-6 hours (13.33%) or more than 12 hours (20%). The latter result is particularly alarming because it shows that construction of experiments is a technology-influenced task that may be greatly facilitated by usage of intelligent multimedia retrieval algorithms and procedures. A stimuli construction process that lasts several hours or days is very demanding and labor intensive for the researcher. It also impairs experiment flexibility and repeatability. For comparison, a document retrieval task using the intStimGen tool would last between several milliseconds and seconds depending on the database size and query complexity. In any case this is negligible compared to the typical duration of a manual stimuli sequence construction.

Construction of stimuli sequences is still almost entirely unsupported by intelligent software tools. Almost half of the all researchers explore multimedia stimuli databases only manually (46.67%), by simultaneously looking over multimedia documents and their descriptions. However, 23.33% of researches retrieve stimuli both manually and with a software tool. Some researchers use general purpose tools like Statistica and SPSS for stimuli retrieval, or their teams have developed custom-made software applications for this purpose (20%).

Most researchers are ambiguous towards practical difficulties in construction of stimuli sequences, yet a 86.67% of them would like to accomplish this task even faster and more efficiently. This answer could imply that researchers have not yet used multimedia stimuli management tools but they do recognize that the purely technical side of their work should be somehow streamlined and made easier.

A clear majority of the participants (73.33%) think that current emotionally annotated databases lack at least some stimuli with a particular semantic or emotional content that could be useful in their work. Also, 93.33% of the participants consider real-life video sequences in emotion elicitation procedures to be useful or very useful. Combined with previous answers this once again accentuates the need to expand the content of multimedia stimuli databases and create new databases with rich and personalized emotion-provoking content that provides numerous benefits in the emotion elicitation experiments [13].

Most participants (60%) deem stimuli descriptors to be at least in some way inadequate, ambiguous or insufficient in conveying the true semantic and emotional content of multimedia stimuli. These problems motivate interdisciplinary research into methods for improving annotation of multimedia stimuli. Formal knowledge representation and automated reasoning techniques can be successfully applied to impair deficiencies of stimuli databases and represent the best option for their optimization. The aforementioned intStimGen application was developed with intention to implement these advanced stimuli retrieval methods and improve construction, utilization and proliferation of multimedia stimuli databases.

Finally, it should be mentioned that a number of participants spontaneously expressed the support for the development of intelligent tools for multimedia stimuli retrieval and wished to remain in contact after the survey was completed and evaluate the tools in their research.

## V. CONLUSION

Although not large by the complexity of posed questions or by the number of participants, this survey may be considered a success. Aggregated answers follow Gaussian distribution and present statistically significant indications of the target population opinions. Motivation of the experts who have taken part in the survey demonstrates the importance of the issues involved, i.e. current trends in usage patterns of multimedia stimuli databases and intelligent tools for their management.

The survey was initially conceived as a test for validity of the intelligent stimuli sequence generator concept, but apart from succeeding in its verification the survey provided an insight in the needs and preferences of the professional community.

The collected results may be considered as an unmistakable empirical indication of the validity of the intelligent stimuli generator concept and provides an encouragement for continuation of the work in this area.

The author of the survey would like to cordially thank all experts who have taken part in the survey.